\documentclass[twocolumn,superscriptaddress,showpacs,preprintnumbers,pra,showkeys]{revtex4}
\usepackage{graphicx}
\usepackage{amsmath,amssymb}
\newcommand{\beq}{\begin{equation}}
\newcommand{\eeq}{\end{equation}}
\newcommand{\beqa}{\begin{eqnarray}}
\newcommand{\eeqan}{\end{eqnarray*}}
\newcommand{\beqan}{\begin{eqnarray*}}
\newcommand{\eeqa}{\end{eqnarray}}
\newcommand{\bra}[1]{\left\langle{#1}\right|}
\newcommand{\ket}[1]{\left|{#1}\right\rangle}
\newcommand{\ip}[1]{\left\langle{#1}\right\rangle}

\newcommand{\eqr}[1]{Eq.~(\ref{#1})}
\newcommand{\sfrac}[2]{{\scriptstyle \frac{#1}{#2}}}

\begin{document}

\title{Quantum protocols for anonymous voting and surveying}
\author{J.A. Vaccaro}
\affiliation{Centre for Quantum Computer Technology, Centre for Quantum Dynamics,
School of Science, Griffith University, Brisbane 4111, Australia}
\affiliation{Quantum Physics Group, STRI, University of
Hertfordshire, College Lane, Hatfield, AL10 9AB, U.K.}
\author{Joseph Spring}
\affiliation{Quantum Information Group, School of Computer Science, University of
Hertfordshire, Hatfield AL10 9AB, Hertfordshire, UK}
\author{Anthony Chefles}
\affiliation{Quantum Information Processing Group, Hewlett-Packard Laboratories,
Filton Road, Stoke Gifford, Bristol BS34 8QZ, UK.}

\date{\today}
\pacs{03.67.-a, 03.67.Dd, 03.65.Ud, 03.65.Vf}
\keywords{voting, quantum data security, quantum information}

\begin{abstract}
We describe quantum protocols for voting and surveying. A key feature of our schemes is the use of
entangled states to ensure that the votes are anonymous and to allow the votes to be tallied. The
entanglement is distributed over separated sites; the physical inaccessibility of any one site is
sufficient to guarantee the anonymity of the votes. The security of these protocols with respect to
various kinds of attack is discussed.  We also discuss classical schemes and show that our quantum
voting protocol represents a $N$-fold reduction in computational complexity, where $N$ is the
number of voters.
\end{abstract}

\maketitle
\section{Introduction}
\label{sec:1}

A well-established consequence of the proven security \cite{QKD_security} of quantum key
distribution is that quantum systems can be used for unconditionally secure classical information
transmission. It is widely believed that classical cryptosystems cannot distribute a key such as a
one-time pad with unconditional security.  Without a one-time pad classical cryptosystems are
secure only on condition that insufficient computational resources are available to render them
vulnerable.  While this assumption is reasonable at the present time, we anticipate that, in the
future, quantum computers will be developed which may be used to attack cryptosystems reliant upon
either integer factorization or discrete logarithm evaluation using, for example, Shor's algorithm.
The security of most common cryptosystems is dependent on the fact that no known efficient
classical algorithm exists that can break particular implementations used within the time period
for which security is desired. As such, although quantum computation will bring many benefits, it
will also be highly disruptive with regard to data security.

For this reason, it is highly probable that, to create cryptographic keys, we will have to turn to
quantum mechanics to provide us with alternative means of establishing security. Fortunately, the
practical implementation of quantum cryptography has advanced considerably over that of quantum
computation.  We are therefore unlikely to face a `security gap'.  Indeed, the first commercial
quantum key distribution systems have recently appeared on the market \cite{Commercial_Suppliers}.
Developments such as these increase our confidence in quantum cryptography and lead us to enquire,
more broadly, about which tasks requiring secure communication could be implemented using quantum
states.

It is known that for some tasks, such as bit commitment, quantum mechanics cannot help
\cite{QuBitCom}.  However, for others, such as secret sharing, a number of novel quantum protocols
have been developed \cite{SecSharing}. In an $(n,k)$ secret sharing scheme, a classical message is
split among $n$ parties.  The key property of such a scheme is that no less than $k$ of these
parties can extract any information about the secret, while any $k$ of them can extract the secret
in its entirety.

In some situations, it is more desirable that the identity of the person who sent the message,
rather than the message itself, be kept secret \cite{Christandl}. Examples include elections,
anonymous ballots and referendums.  Here each voter should should feel able to cast their vote
without the prospect of coercion or repercussion.  Only collective features of the set of votes,
such as the tally of `yes' and `no' votes, are calculated and made public.

In this paper, we describe novel quantum protocols for voting and a related task that we term
surveying.  Surveying is similar to voting in most respects.  The main difference is that in
surveying, the value of the vote cast is not restricted to a binary `yes' or `no' but may take any
integer value.  As such, surveying corresponds to collecting estimates of some numerical quantity,
such as profit and loss values. The identities of the people who make each bid are kept private,
although the sum of the bids is made public. We also analyze the security of these protocols under
some simple attacks.

To set our work in context, we review in section \ref{sec:2} a selection of protocols currently
employed in secure classical election schemes. We devote section \ref{sec:3} to the description of
our protocols. The first of these is a simple quantum protocol for {\it comparative voting.} Here,
we consider two parties voting on a question with a `yes' or `no' answer.  The aim is not to
determine the tally itself, but to determine whether or not both parties voted identically without
knowing the value of each of the votes. We show that this is possible by encoding the voting
information in an entangled state.

Subsequently, we describe a protocol for {\it anonymous surveying}. It proves robust against
certain kinds of attack. We discuss adaptations of the protocol for an {\it anonymous ballot} for
binary-valued ballots and the relationship between the privacy of a vote and the ability for a
voter to cheat by making multiple votes. We conclude in section \ref{sec:4} with a discussion of
our results.

\section{Classical voting protocols}
\label{sec:2}

Various properties have emerged from the literature as being desirable attributes of classical
secret ballot voting schemes. Amongst these is the concept of {\it resilience} which involves the
properties of universal verifiability, privacy, and robustness ~\cite{CFSY:1996}.  A {\it
universally verifiable} election scheme is a scheme deemed open to scrutiny by all interested
parties. Compliance with this property ensures that ballots are carried out correctly and that
subsequent tallies are fairly assessed.  For a scheme satisfying the {\it privacy} property an
honest participant is assured that their vote remains confidential, provided that the number of
attackers does not grow too large.  With the property of {\it robustness}, an election scheme has
the capacity to recover from faults again, provided that the number of parties involved does not
grow too large. Schemes satisfying these three properties are said to be {\it resilient}. Another
desirable property of an election scheme, particularly as a counter to the risk of vote buying or
coercion, is that it is {\it receipt-free}. Receipt-free election schemes ensure that voters cannot
prove, to other parties, the particular vote cast within the scheme ~\cite{BT:1994, HS:2000}.
Further `desirable properties' are to be found in the literature, for example ~\cite{Sch:1996}.

Voting protocols performed within a classical setting are in general grouped according to their use
of homomorphisms, MIX nets and blind signatures.

{\em Homomorphic election schemes}.  These
~\cite{Ben:1987,BY:1986,CF:1985,CGS:1997,CFSY:1996,SK:1994} involve the use of a homomorphic,
probabilistic encryption scheme consisting of a plaintext space $\mathcal{V}$, a ciphertext space
$\mathcal{C}$ (each of which form  group structures ($\mathcal{V}, \circ$) and ($\mathcal{C},
{\circ}'$) under appropriate binary operations $\circ$ and ${\circ}'$) together with a family of
homomorphic encryption schemes \{$E_i \}_{i \, \in \, \mathbb{N}^+}$ such that
$E_i:\mathcal{V}\longrightarrow \mathcal{C}$ by $v \mapsto c = E_i(v)$. The homomorphic property
~\cite{CGS:1997} may be defined as follows: let $c_j = E_{i_j}(v_j)$ and $c_k = E_{i_k}(v_k)$ for
$j, {i_j}, k, {i_k} \in \mathbb{N}^+$; then $\exists i \in \mathbb{N}^+$ s.t. $c_j {\circ}' c_k =
E_i (v_j \circ v_k )$. Homomorphic election schemes are important since they allow one to derive
tallies without the need to decrypt  individual votes.  Such schemes lead to resilient election
schemes ~\cite{CGS:1997,CFSY:1996}.

{\em MIX net schemes.}  MIX nets were first introduced by Chaum ~\cite{Ch:1981}, and have found
applications in scenarios involving anonymity, elections and payments.  A MIX net election scheme
involves the use of `shuffle machine agents' referred to as MIX servers, ~\cite{PIK:1993} which
take as input a ciphertext vector (these could be for example, encrypted votes) {$(c_1, c_2, ... ,
c_n )$} $\in \bigoplus_{i = 1}^{n} \mathcal{C}_i$ submitted by for example \lq voters',  $(v_1 ,
... , v_n) $ and produces as output a permuted vector (in which the components are shuffled) of
corresponding output (for example, decrypted votes ) such that the link between the source for each
ciphertext (`encrypted vote') and its resulting plaintext ('vote') remains hidden. The resilience
properties of privacy, verifiability and robustness may be presented in terms of `$t$-privacy',
`$t$-verifiability' and `$t$-robustness', where it is understood that $t$ refers to the number of
malicious MIX servers that the scheme can withstand given at most $n - 2$ malicious sources.  A
scheme satisfying the above three $t$-properties is said to be $t$-resilient ~\cite{DK:2000}.

The development of classical MIX net schemes to achieve, in particular, privacy initially led to
ciphertext whose size was proportional to the number of MIX servers involved in the scheme. This
problem was resolved by Park, Itoh and Kurosawa ~\cite{PIK:1993}, resulting in ciphertext whose
length was independent of the number of MIX servers. Sako and Kilian ~\cite{SK:1995}, produced a
general MIX net scheme satisfying verifiability but failing with regard to robustness. The first
resilient MIX net scheme was produced by Ogata, Kurosawa, Sako and Takatani
~\cite{OKST:1997,DK:2000}.

{\em Blind signature schemes.}  These were also introduced by
 Chaum ~\cite{Ch:1983}, and have been developed with
applications in anonymity, election and payment schemes.  The basic concept involves obtaining a
signature to authenticate a `message', for example an encrypted vote, without the signer being able
to observe the message (`vote') itself or its signature. Verification regarding the signature is
however supported by such schemes whilst maintaining privacy regarding the actual plaintext.  A
signer is thus denied the ability to link a particular plaintext with its corresponding `blind'
signature ~\cite{CPS:1994}. Variations upon such schemes are to be found with for example `Fair
Blind Signatures' ~\cite{SPC:1995} in which the possibility of, for example, blackmail is discussed
~\cite{SN:1992}.

{\em  Sender untraceability schemes.}  These schemes allow information to be sent anonymously. For
example, in Chaum's Dining Cryptographers' Problem ~\cite{Ch:1988Diner} a group of diners wish to
determine if either an external agency or one of the group is paying anonymously for the meal. The
solution requires 1 bit of information to be broadcast anonymously using a communication channel
available to all diners. The simplest situation occurs for three diners with only two possible
scenarios: one diner is to pay the bill or no diners pay the bill. The diner who pays broadcasts
the message 1 in the following way. Each diner shares a single binary-digit one-time pad with the
other two. The broadcast is executed by each diner adding the two numbers on the one time pads he
or she holds. If one of the diners is paying he or she adds 1 to the value of the sum. The results
modulo 2 are announced publicly to all diners. The sum of the 3 broadcast messages modulo 2 is 1
only if the message 1 is sent by a paying diner otherwise it is 0. Thus a message is broadcast but
the identity of a paying diner is untraceable.

The security of a classical scheme is deemed to be one of two varieties: computational or
unconditional (also known as information-theoretic) security \cite{Mau:1999}.  A scheme which can
be broken in principle but requires more computing power than a realistic adversary can access in a
given critical time is deemed computationally secure.  Examples are schemes based on the integer
factorization problem and the discrete logarithm problem.  Such computationally secure schemes are
under threat from quantum computing. On the other hand, a scheme which is secure even if an
adversary has unlimited computing resources is said to be unconditionally secure. A one time pad
encryption scheme is unconditionally secure. Homomorphic maps and mixed nets not based on the one
time pad are computationally secure. Blind signatures can be applied in an unconditionally secure
manner to authenticate a vote and sender untraceability provides anonymity with unconditional
security. Chaum's secret ballot protocol \cite{Chaum:1988Ballot}, which uses blind signature and
sender untraceability schemes, allows unconditionally secret voting. The sender untraceability
component of the protocol requires one-time pads between all pairs of voters, that is $N(N-1)/2$
one time pads are required for a ballot with $N$ voters.

\section{Quantum protocols for anonymous surveying and voting}
\label{sec:3}

In this paper we examine a number of quantum protocols for ballots \cite{NoteNewPapers, HZBB}. In
light of the foregoing classical schemes, we desire the ballots to satisfy the following general
rules:
\begin{description}
\renewcommand{\theenumi}{(R\arabic{enumi})}
\item[\rm(R1)]
  \label{secrecy}
  The vote of each voter should be kept
  secret from all other voters.
\item[\rm(R2)]
  \label{tally}
  The person (the tallyman) calculating the collective quantity
  should not be able to gain information about the voting of
  individual voters.
\item[\rm(R3)]
  \label{receipt}
  The votes should be receipt-free. This is to say that it should
  impossible for a voter to prove how they voted to a third party,
  even if they wanted to.  This condition thwarts vote buying and
  ensures the uncoercibility of the voter.
\end{description}

\noindent An additional rule applies for the special case of a restricted ballot where the range of
values of each vote is restricted:
\begin{description}
\renewcommand{\theenumi}{R\arabic{enumi}}
\setcounter{enumi}{3}
\item[\rm(R4)]
  \label{restrict}
  A voter may not make more than one vote, that is, the value of
  each vote should not count as more than one vote.
\end{description}
We call a ballot where the votes are restricted to being binary-valued a {\em binary-valued
ballot}. A specific example is a simple referendum. We call a ballot that satisfies the first three
rules but not the fourth an {\em anonymous survey}.

Additional (ancillary) people may be involved in the ballot, but they must not have access to any
more information about the voting than the tallyman. There are a number of different kinds of
ballots depending on the nature of the ballot question and the collective voting information
required.

\subsection{Comparative ballot}
The first quantum voting protocol we shall describe is a simple comparative protocol which we call
a {\em comparative ballot}. Consider two voters, Alice and Bob, voting on a question with a
response of either `yes' or `no. There is also a tallyman, whose principal aim is to determine
whether or not they agree, i.e., whether or not they have cast the same vote.

We wish the result of this comparison task to be deterministic and always correct (i.e.
unambiguous).  It should be noted in this context that, if instead of classical information, we
wish to unambiguously compare pure quantum states, then certain restrictions would apply. In
particular, the possible states would have to be linearly independent \cite{CAJ}.

Alice and Bob are assumed to be at spatially separated sites $A$ and $B$. The protocol entails
beginning with the {\em ballot state} representing one particle shared between the two sites $A$
and $B$:
\beq \ket{C_{0}}=\frac{1}{\sqrt{2}}(\ket{1,0}+\ket{0,1})\ .
      \label{ballot_1}
\eeq
Here, $\ket{n,m}\equiv\ket{n}_{A}\otimes\ket{m}_{B}$ represents $n$ particles (bosons) occupying a
spatial mode at site $A$ and $m$ particles occupying an orthogonal spatial mode at site $B$ in
second quantization notation. A voter makes a `yes' vote by applying the operator
$\exp(i\hat{N}\pi)$, where $\hat{N}$ is the voter's local particle number operator and
$\exp(i\hat{N}\pi)\ket{n}=\exp(in\pi)\ket{n}$. A `no' vote is cast by simply doing nothing, which
formally amounts to applying the identity operator. If both voters make the same vote, then the
ballot state is unchanged (up to a possible overall sign inversion.) If, on the other hand, their
votes are different, then the ballot state is transformed into
\beq
      \ket{C_{1}}=\pm\frac{1}{\sqrt{2}}(\ket{1,0}-\ket{0,1})\ ,
      \label{ballot_2}
\eeq
where the sign $\pm$ depends on who votes `yes'.  At all times, the voter at one site cannot
determine the vote cast at the other site. This is because the reduced density operator
representing the state of the particle at each site is always the maximally-mixed state
$(\ket{0}\bra{0}+\ket{1}\bra{1})/2$.  The voting is kept strictly private to the respective voters.

The two-particle state is then transferred to the site of the tallyman who performs a measurement
in the basis $(\ket{1,0}\pm\ket{0,1})\sqrt{2}$.  The tallyman is able to discern whether the voters
have made the same or opposite vote even though he is unable to determine how each voter cast their
vote. This situation is similar in some respects to the Deutsch-Jozsa algorithm for deciding the
balanced or unbalanced nature of a binary function \cite{deutsch}. Also the single particle state
in \eqr{ballot_1} is the same state used in the data hiding protocol of Verstraete and Cirac
\cite{VC}. Whereas in Ref. \cite{VC} a third party stores a secret in the state shared by Alice and
Bob, here it is Alice and Bob who store secrets in the shared state. Our protocol satisfies rules 
1 and 2 of private ballots, namely each vote is is known only to its corresponding voter, and the 
tallyman has access only to the collective (comparative) information. The potential for cheating 
is limited by the very nature of the comparative ballot; voters can only make a single vote.  The 
application of an operator other than $\exp(i\hat{N}\pi)$ is interpreted as indecision in the 
sense that the result of repeated identical ballots is stochastic.

\subsection{Anonymous survey}

In the above protocol, the voting information was stored in locally inaccessible phase factors in a
entangled state.  This technique can be applied to other situations where we wish to maintain
anonymity.  One such scenario is as follows.

Let us suppose that the chief executive officer (CEO) of a firm wants to gauge the effect of a
possible action; he surveys the opinion of his management team to find out what each member thinks
the likely profit (or loss) will be.  To avoid the dishonest responses due to rivalry, grovelling,
fear of repercussions etc., the CEO wants the survey to be anonymous. The managers must report the
estimated profit or loss for their particular department. The CEO is interested in the total for
the whole company. An alternative, but essentially equivalent, situation is that the managers
estimate the total profit or loss for the firm as a whole, and the CEO wants the average of the
estimates, that is, the sum of the estimates divided by the number of managers. In both cases the
sum of the estimates is made public and the individual amounts are private.  We call the protocol
for determining the sum while keeping the individual amounts secret an {\em anonymous survey}.

An anonymous survey should obey the rules R1-R3. We now describe a quantum protocol that satisfies 
these rules. We retain the general terminology of `voters', `votes' and the `tally'. The basic 
principle of the protocol involves a {\em two-mode} discrete phase state \cite{PB, KitYam} shared 
between the voters and the tallyman. A vote is made by translating the phase value of the phase 
state. Due to the shared nature of the state, the actual value of the phase is hidden from both 
the voters and the tallyman in a manner analogous to secret sharing \cite{VC}.

We again use a system of identical particles in the second quantization formalism.  We employ $N$
particles, where $N$ is equal to or larger than the number of voters.  The particles are prepared
in the following {\em ballot state}:
\beq
  \ket{B_0} = \frac{1}{\sqrt{N+1}}\sum_{n=0}^N \ket{N-n,n}\ ,
  \label{ball_state}
\eeq
where $\ket{n,m}\equiv\ket{n}_T\otimes\ket{m}_V$ and $\ket{n}_V$ ( $\ket{n}_T$) represents $n$
particles localized in a spatial mode at site $V$ (respectively $T$). The sites $V$ and $T$ are
assumed to be remote from each other. Voters have access only to $V$ and not $T$, whereas the
tallyman has access only to $T$ and not $V$. Voter $i$ makes a vote by applying the phase shifting
operation $\exp(i\hat{N}_V\delta_i)$ to the spatial mode at site $V$, where
$\hat{N}_V\ket{n}_V=n\ket{n}_V$ and $\delta_i=\nu_i\pi/(N+1)$ for a vote corresponding to an amount
$\nu_i\ $. For example, after the vote of the first voter the ballot state becomes:
\beq
  \ket{B_1} = \frac{1}{\sqrt{N+1}}\sum_{n=0}^N \exp(in\delta_1)\ket{N-n,n}\ .
    \label{1st_vote}
\eeq
The second voter makes a vote in a similar manner with the phase shifting angle $\delta_2\ $. This
voting process is repeated for all voters. The resulting ballot state after the $m$th voter is
\beq
  \ket{B_m} = \frac{1}{\sqrt{N+1}}\sum_{n=0}^N
  \exp(in\Delta_m)\ket{N-n,n}\ ,
    \label{many_votes}
\eeq where $\Delta_m=\sum_{i=1}^m\delta_i\ $.  The net value of
the accumulated votes is $M_{m}=\sum_{i=1}^m\nu_i$ which can be deduced from the final phase angle
using
\beq
  \Delta_m=\frac{2{\pi}M_{m}}{N+1}\ .
  \label{tally_M}
\eeq

Note that at any point in this operation the tallyman, who does not have access to the site $V$,
can only see the mixed state
\beq
   {\rm Tr}_V (\ket{B_m}\bra{B_m})=\frac{1}{N+1}\sum_{n=0}^N
   (\ket{n}\bra{n})_T\ ,
\eeq
which is invariant under the phase shifting operation. Likewise, the voters, who do not have access
to the site $T$, can only see the mixed state
\beq
   {\rm Tr}_T (\ket{B_m}\bra{B_m})=\frac{1}{N+1}\sum_{n=0}^N
   (\ket{n}\bra{n})_V\ .
\eeq
The voting of individual voters is therefore secret from other voters and the tallyman.

At the end of the survey the particles at site $V$ are translated to a mode at site $T$ so that the
ballot state is as in \eqr{many_votes} but with $\ket{n,m}\equiv\ket{n}_T\otimes\ket{m}_{T}$. We
imagine that the tallyman has access to both modes at site $T$. The states $\ket{N-n,n}$ form an
orthonormal basis for an $N+1$ dimensional subspace.  We define another orthonormal basis as
follows \cite{PB}:
\beq
   \ket{T_n}=\frac{1}{\sqrt{N+1}}\sum_{k=0}^N
   \exp(ink\theta)\ket{N-k,k}\ ,
\eeq
where $\theta=2\pi/(N+1)$ and
\beq
   \ip{T_n|T_m}=\delta_{nm}\ .
\eeq
The ballot states are all eigenstates of the {\em tally operator}:
\beq
   \hat{T}=\sum_{n=0}^N n \ket{T_n}\bra{T_n}\ .
\eeq
To find the tally, the tallyman finds the expectation value of the tally operator which yields:
\beq
    \bra{B_m}\hat{T}\ket{B_m}=M_{m}\ .
\eeq
The tallyman can access the tally only once he is in possession of all particles.  The voting of
individual voters is kept secret from both the tallyman and the voters while the particles are
shared between the sites.\\

{\em Attack by colluding voters.} Two voters, $A$ and $B$, can collude in the following manner to
deduce information about other voters. First we write the ballot state as
\beq
  \label{collusion}
   \ket{B_0}=\frac{\sqrt{N+1}}{2{\pi}}\int_{0}^{2{\pi}}
   \ket{\psi(\theta)}\ket{\phi(\theta)}d{\theta}\ ,
\eeq
where the single-mode phase states \cite{PB} are given by
\beqa
   \ket{\psi(\theta)}&=&\frac{1}{\sqrt{N+1}}\sum_{n=0}^{N}e^{-in\theta}\ket{N-n},\\
   \ket{\phi(\theta)}&=&\frac{1}{\sqrt{N+1}}\sum_{n=0}^{N}e^{in\theta}\ket{n}\ .
\eeqa
Imagine that voter $A$ locally measures the phase of the system at the voting site; this will
project the system onto a state of the form $\ket{\psi(\theta')}\ket{\phi(\theta')}$, where
$\theta'$ represents the outcome of the measurement.  The tally $M$ of votes of subsequent voters
then accumulates locally in the phase of the local system, resulting in the state
$\ket{\psi(\theta')}\ket{\phi(\theta'+M\delta)}$. Subsequent measurement of the phase by voter $B$
and comparison with the phase measured by $A$ will then reveal the amount
$M$.\\

{\em Detection of attack.} We note that the ballot state, in the absence of the attack, has a fixed
number of particles $N$.  In contrast the projection onto a phase state induces a distribution of
particles, and so the attack alters the total particle number on average. Thus the attack can be
detected by the tallyman making a measurement of the total particle number $\hat N$ and checking if
it differs from $N$.  The probability of detecting the attack by this method is $1-P_N$ where
$P_N=1/(N+1)$ is the probability of finding $N$ particles in the
state $\ket{\psi(\theta')}\ket{\phi(\theta')}$.\\

{\em Defence.}  A defence against this colluding attack is to use a multiparty ballot state such as
\beq
  \ket{B_0^\prime} = \frac{1}{\sqrt{N+1}}\sum_{n=0}^N
  \ket{K(N-n),n,n,\cdots,n}\ ,
  \label{ball_state_multi}
\eeq
where $\ket{i,j,\cdots,k}\equiv\ket{i}_T \otimes\ket{j}_{V_1}\otimes\cdots\ket{k}_{V_K}$ for $K$
voting sites $V_i$, $i=1,\cdots,K$ where $K$ is equal to or larger than the number of voters. Each
voter $i$ is assigned a unique voting site $V_i$ for casting a vote as before, i.e. using the phase
shifting operation $\exp(i\hat{N}_{V_i}\delta_i)$ to the spatial mode at site $V_i$. The colluding
attack is foiled because each voting site is used by a single voter. The final multipartite ballot
state is
\beq
  \ket{B_m^\prime} = \frac{1}{\sqrt{N+1}}\sum_{n=0}^N
  \exp(in\Delta_m)\ket{K(N-n),n,n,\cdots,n}
\eeq
and the corresponding multipartite tally operator is given by
\beq
   \hat{T}^\prime=\sum_{n=0}^N n \ket{T_n^\prime}\bra{T_n^\prime}\ ,
\eeq
where
\beq
   \ket{T_n^\prime}=\frac{1}{\sqrt{N+1}}
   \sum_{k=0}^N \exp(ink\theta)\ket{K(N-k),k,k,\cdots,k}\ .
\eeq
After all the particles are translated to the tallyman, the tallyman determines the value of the
tally from the expectation $\bra{B_m^\prime}\hat{T^\prime}\ket{B_m^\prime}=M_m$ in the same manner
as before.

\subsection{Anonymous binary-valued ballot}

A special case of an anonymous survey is an anonymous ballot for binary-valued votes which we call
an {\em anonymous binary-valued ballot}. A simple referendum would be a specific example of this
kind of ballot. Here, instead of votes being an arbitrary integer, each vote is of a binary nature,
`yes' or `no', corresponding to an answer to a public question. The anonymous survey protocol above
could be used for an anonymous binary-valued ballot provided the voters were honest and restricted
their vote value accordingly. For example, voter $k$ could choose a phase shift angle of
$\delta_i=0$ for a `no' vote and $\delta_i=2\pi/(N+1)$ for a `yes' vote.  The tally $M_m$ in
\eqr{tally_M} then corresponds to the number of `yes' votes (the number of `no' votes being
calculated from the number of participating voters less $M_m$). There are special situations where
it is in the voters' interest to vote honestly, for instance, where the public ballot question is
one of a personal nature (requiring anonymity) and where the voters want to know the true
proportion of voting population sharing the same view.  Of course, in general, the voters may be
tempted to vote more than once and, for example, a voter may choose $\delta_k=4\pi/(N+1)$ to record
two `yes' votes. It is therefore important that rule R4 is enforced in anonymous binary-valued 
ballots.

The underlying reasons why the anonymous protocol does not satisfy R4 are rather simple and quite 
general. Consider two parties Alice and Bob, and the entire initial ballot state 
$|B_{0}{\rangle}$. Let $Y_{A}$ and $Y_{B}$ be the unitary operators used by Alice and Bob 
respectively to register `yes' votes. It is assumed that they would both vote `no' by applying the 
identity operator.  Any tensor product structure arising from, e.g., Alice and Bob registering 
their votes on different systems is taken to be implicit in these definitions. Let us now consider 
the implications of anonymity and define
\begin{equation}
{\Omega}=||(Y_{A}-Y_{B})|B_{0}{\rangle}||,
\end{equation}
where $|||{\psi}{\rangle}||\equiv{\langle}{\psi}|{\psi}{\rangle}$. Anonymity requires that for
either a `yes' or `no' vote, it should be impossible to determine who made this vote.  It therefore
requires that ${\Omega}=0$.

Let us now suppose that one of the parties, say Alice, wishes to cheat. She wishes to vote `yes'
twice by making it appear as though both she and Bob have voted `yes', where Bob is some other
voter who has voted `no' (and in a sufficiently large ballot that such a party exists is a fair
assumption).  Suppose that Alice votes after Bob.  Then, for Alice's cheating to go undetected, the
states $Y_{B}Y_{A}|B_{0}{\rangle}$ and $Y_{B}^{2}|B_{0}{\rangle}$ must be completely
indistinguishable. That this is the case under conditions of anonymity can be seen from
\begin{equation}
  ||(Y_{B}Y_{A}-Y_{B}^{2})|B_{0}{\rangle}||
  =||Y_{B}(Y_{A}-Y_{B})|B_{0}{\rangle}||={\Omega}\ ,
\end{equation}
where the last step follows from the unitary invariance of the norm.  Under conditions of
anonymity, ${\Omega}=0$ and so the two states are completely indistinguishable, concealing Alice's
actions.

We may also require that Alice can cheat undetected irrespective of the order in which she and Bob
cast their votes.  This will be the case if
\begin{equation}
[Y_{A},Y_{B}]|B_{0}{\rangle}=0\ .
\end{equation}
This condition is automatically satisfied if Alice and Bob register their votes on different
systems, or the same operators if they use a common system, as is the case in the protocol we have
described.

It follows that to protect against this kind of cheating strategy, some of the properties of the
protocol must be changed.  The key one seems to be unitarity.  It is therefore interesting to
explore the possibility of maintaining anonymity yet preventing cheating if we drop unitarity and
use irreversible operations to register votes.  We will now see that doing so does allow for a
certain degree of improvement.  In particular, we will see how the use of irreversibility can limit
the extent of one party's cheating to a mere 0.5 votes, and only at the expense of reduced privacy,
in contrast with the limitless extent to which they can cheat using unitary operations with
impunity.

The possibility of cheating can be restricted by introducing an element of irreversibility into the
voting procedure.  One way of doing this would be to restrict the operation able to be performed by
each voter to the appropriate values by directly restricting the macroscopic devices used to
perform the voting operations. However, since the restricted device is not in the total control of
the voter (by necessity) it seems likely that evidence of the action taken by the voter could be
traced in the local environment and so criterion R1 is not guaranteed to be satisfied. This is 
clearly a general problem with irreversible operations.

One way to restrict the votes without macroscopic means is to replace the initial ballot state
\eqr{ball_state} with
\beq
  \ket{B_0^{\prime\prime}}
  = \frac{1}{\sqrt{N+1}}\sum_{n=0}^{N} \ket{2(N-n),n,n}\ ,
  \label{2agents}
\eeq
where $\ket{i,j,k}\equiv\ket{i}_T\otimes\ket{j}_{V_1}\otimes\ket{k}_{V_2}$ and $V_1$ and $V_2$
label two voting sites which are controlled by two ballot agents $A_1$ and $A_2$, respectively.
Each voter privately records their vote in a (separate) pair of qutrits (i.e. a pair of spin-1
systems) with the state $(\ket{0,-1}+\ket{-1,0})/\sqrt{2}$ for a `no' vote or the state
$(\ket{0,1}+\ket{1,0})/\sqrt{2}$ for a `yes' vote. Here the qutrit states $\ket{-1}$, $\ket{0}$ and
$\ket{1}$ correspond to eigenstates of the $z$ component of spin. One qutrit is given to each of
the ballot agents who locally apply the operation
\beq
   \exp[i\hat N^{(1)} \delta (\sfrac{1}{4}+\sfrac{1}{2}\hat\sigma_z^{(1)})]
   \otimes
   \exp[i\hat N^{(2)} \delta
   (\sfrac{1}{4}+\sfrac{1}{2}\hat\sigma_z^{(2)})]\ ,
   \label{Agent_Operation}
\eeq
where $\delta=2\pi/(N+1)$, $\sigma_z^{(i)}$ is the $z$ component of spin for qutrit $i$ at site
$V_i\ $, and $\hat N^{(i)}$ operates on the spatial mode in $\ket{B_0^{\prime\prime}}$ at site
$V_i$. This allows the votes to accumulate in the discrete phase angle as before and ensures that
each voter casts a single vote.

{\em Attacks by ballot agents.} The 2 separated ballot agents cannot separately learn the nature of
the vote with certainty. For example, one ballot agent may measure the $z$ component of spin on the
qutrit at his site.  This would reveal the nature of the vote only {\em half} of the time.

{\em Defence: tamper evidence.} The attack by the ballot agents will result in the qutrit pair
being in a product of eigenstates of the $z$ component of spin. Thus to detect the attack, the
qutrit system should be immediately returned to the voter following the action by the ballot
agents, and be subjected to measurement in a basis which includes the states representing `yes' and
`no' votes to determine if the state has been changed. The attack will be detected one half of the
time. Any instances of changed states are made public, and hence attempted cheating by a ballot
agent will be detected, on average. The qutrit system is therefore {\em tamper evident}, on
average, in this sense.

Alternatively, the privacy of the vote can be increased by increasing the number of ballot agents
and the number of qutrits used to store each vote. For example, in a 3-qutrit, 3-ballot agent
scheme, the ballot state would be
\beq
  \ket{B_0^{\prime\prime\prime}} = \frac{1}{\sqrt{N+1}}\sum_{n=0}^{N}
  \ket{3(N-n),n,n,n}\ .
  \label{3agents}
\eeq
The votes would be made by preparing the states $(\ket{0,0,-1}+\ket{0,-1,0}+\ket{-1,0,0})/\sqrt{3}$
for a `no' vote or the state $(\ket{0,0,1}+\ket{0,1,0}+\ket{1,0,0})/\sqrt{3}$ for a `yes' vote, and
each ballot agent $A_i$ applies the operation \beq
   \exp[i\hat N^{(i)} \delta (\sfrac{1}{6}+\sfrac{1}{2}\hat\sigma_z^{(i)})]
\eeq
to their local qutrit system and the spatial mode of $\ket{B_0^{\prime\prime\prime}}$. A
measurement of the $z$-component of spin of one of the qutrits by a ballot agent will now reveal
the value of the vote only {\em one third} of the time,
making the vote more private.\\

{\em Attacks by voters.}  A voter need not prepare the states corresponding to a 'yes' or 'no'
vote.  Indeed, he may try to maximize the value of his vote, for example, by preparing the qutrit
pair in the {\em cheat state} $\ket{1,1}$.  The action of the operator in Eq.\
(\ref{Agent_Operation}) applied by the two ballot agents then increases the phase angle of the
ballot state by $3\delta/2$, that is, by 1.5 votes \cite{Tradeoff}.
\\

{\em Defence: tamper-evidence versus vote-value tradeoff.}  The attack by the voter may increase
the value of the vote, but this is at the expense of abandoning the tamper-evident nature of the
qutrit system. A measurement of the $z$ component of spin of one of the qutrits in the cheat state
will reveal the value of the vote without altering the state; this allows a ballot agent to
determine the vote without being detected. Hence, a voter may cheat by a half vote but only at the
expense of losing the privacy of his vote.

\section{Discussion}
\label{sec:4}

We have introduced quantum protocols for ensuring the anonymous voting in a number of different
scenarios. Central to the protocols is the ballot state, which is an entangled state shared between
at least two sites. The ballot state stores the tally of the votes which are registered using local
operations. At all times the value of the vote tally stored in the ballot state is not available at
any one site, but only in the collection of the sites. The ballot therefore represents a {\em
distributed memory} and this property ensures the privacy of each vote and thus the anonymity of
each voter. After all votes have been made, the vote tally can be determined by a collective
measurement.

We identified 4 rules (R1-R4) which lead to desirable properties of anonymous ballots.  The 
different kinds of ballot depend on the ballot question, the collective information required and 
the subset of rules.  We described protocols for a comparative ballot, where the `yes' or `no' 
answers of two voters are compared, an anonymous survey, where the voters make anonymous votes of 
integer value, and an anonymous binary-valued ballot, where the `yes' and `no' answers to a ballot 
question are tallied anonymously. We note that an anonymous binary-valued ballot is closely 
related to an election, and indeed, it corresponds to an election for the special case of two 
candidates. We are currently exploring other possibilities.

We have analyzed our protocols against a number of attacks. In particular, we found that for the
anonymous binary-valued ballot the cheating by a voter is associated with reduced security. In one
variant of the protocol, while the value of the vote of an honest voter is 1, a dishonest voter can
make a vote of value equal to 1.5 and essentially cheat by 0.5 vote. However, this occurs at the
cost of reduced privacy since the {\em tamper-evident} nature of the protocol, which allows the
voter to detect an attack by a ballot agent, is only available if an honest vote is made. The
various schemes are unconditionally secure {\it on average} in the sense that attacks can be
detected with non-zero probability.

For comparison, we note that Chaum's classical secret ballot protocol is also unconditionally
secure \cite{Chaum:1988Ballot}.  The secrecy is protected through the use of one-time pads which
are shared between all pairs of voters. In a ballot with $N$ voters, this requires each voter to
distribute $N-1$ one time pads with the other voters.  Thus the computational complexity of the
protocol from the perspective of a voter is of order $N$.  In contract, in our quantum voting
scheme each voter needs to share a {\it single} multiparty entangled state with the tallyman. For
example the tallyman could locally prepare a set of spatial modes in the ballot state and then
teleport the states of the entangled modes to each corresponding voter. The computational
complexity from the perspective of a voter is therefore of order unity, which is an $N$-fold
reduction.

We also found a related property \cite{Tradeoff} for our anonymous binary-valued ballot protocol as
follows: as the protocol is modified to increase the privacy of the vote, the restriction on the
possible values of an individual vote weakens. This appears to be a general trade-off property and
we are currently exploring it in more detail. Indeed our results represent an initial study of this
topic and are not intended to be complete. We hope our results will stimulate further research into
this area.

\section{Acknowledgements}
\label{sec:5}

This project was supported by the EU Thematic Network QUPRODIS.   J.A.V. thanks Dr. Arkadiusz
Or{\l}owski and members of the {\it Sydney Quantum Information Theory Workshop}, February 2006 for
helpful discussions and was supported by the Leverhulme Foundation, the Australian Research Council
and the State of Queensland. A.C. was supported by the Science Foundation Ireland and the EU
project QAP.


\begin{thebibliography}{99}
\bibitem{Tradeoff}
There is a relationship between the privacy of the vote and the bound on the vote. We have already
shown that in the 2-qutrit, 2-ballot agent scheme, the value of an honest vote can be revealed by
an a ballot-agent attack in $1/2$ of the attacks, and the voter can cheat by preparing the state
$\ket{1,1}$ which results in a vote of value 2. In general, in the $n$-qutrit, $n$-ballot agent
scheme, an honest vote can be revealed in $1/n$ ballot-agent attacks and a voter can make a vote of
value $(n+1)/2$.  That is, as $n$ increases, the privacy of the vote increases but the ability to
cheat also increases.
\end{thebibliography}
\end{document}